\documentclass[prl,aps,showpacs,twocolumn]{revtex4}

\usepackage{graphicx}% Include figure files
\usepackage{dcolumn}% Align table columns on decimal point
\usepackage{bm}% bold math
\usepackage{amsmath, amsthm,times}
\usepackage{epsfig}
\usepackage{amssymb}
\usepackage{txfonts}

\begin{document}

%\preprint{LOQCGA_V}%

%\title{Annealing Genetic Algorithm to Optimize Quantum Logic Gates}%
\title{Optimizing Optical Quantum Logic Gates using Genetic Algorithms}%

\author{Zhanghan Wu$^1$}
\author{Sean D. Huver$^1$}%
  \email{huver@phys.lsu.edu}
\author{Dmitry Uskov$^2$}%
\author{Hwang Lee$^1$}%
\author{Jonathan P. Dowling$^1$}%
\affiliation{$^1$Hearne Institute for Theoretical Physics, Department of Physics and Astronomy, Louisiana State
University, Baton Rouge, LA 70803 \\
$^2$Department of Physics and Astronomy, Tulane University, New Orleans, Louisiana 70118}%

\date{\today, V1.1}% It is always \today, today, + version No.

\begin{abstract}
%Abstract here
We introduce the method of using an annealing genetic algorithm to the numerically complex problem of looking for quantum logic gates which simultaneously have highest fidelity and highest success probability. We first use the linear optical quantum nonlinear sign (NS) gate as an example to illustrate the efficiency of this method. We show that by appropriately choosing the annealing parameters, we can reach the theoretical maximum success probability (1/4 for NS) for each attempt. We then examine the controlled-z (CZ) gate as the first new problem to be solved. We show results that agree with the highest known maximum success probability for a CZ gate (2/27) while maintaining a fidelity of 0.9997. Since the purpose of our algorithm is to optimize a unitary matrix for quantum transformations, it could easily be applied to other areas of interest such as quantum optics and quantum sensors. 
\end{abstract}

                              %displaEq:projy desired
\maketitle
%\section{}%Introduction
%section 1: Background on Optimization of LOQSG
Linear optics is an attractive candidate for building quantum computers in large part due to Knill, Laflamme and Milburn \cite{KLM01} and their scheme for non-deterministic quantum gates with projective measurement. The scheme provides a way to build elementary quantum gates with only linear optical elements. The trade off in this scheme is that we can only construct the gates with a certain success probability, i.e., they are non-deterministic. Therefore, one of the major tasks for this program is to figure out a general way to determine the design of gates which possess the highest success probability.

Any quantum gate or linear optical quantum state generator (LOQSG) can be viewed as a unitary transformation which transfers certain input states into desired output states. The goal of designing a LOQSG is to find a proper unitary matrix whose elements can then be implemented with linear optical devices \cite{Reck94, Kok07}. In this paper, we introduce genetic algorithms with a simulated annealing to the problem of optimizing a LOQSG. We first restate the problem so that it is suitable for genetic algorithms. We then briefly discuss the method of genetic algorithms with simulated annealing and a tunable control constraint. Using this method, we first test the efficiency of the algorithm with the nonlinear sign (NS) gate. We choose the NS gate due to its maximum success probability without feedforward having been theoretically proven to be 1/4 \cite{ScheelNS04, EisertNS05}. We then investigate the CZ gate and try to obtain the global maximum for success probability while maintaining a high level of fidelity.

\begin{figure}
\includegraphics[width=8cm]{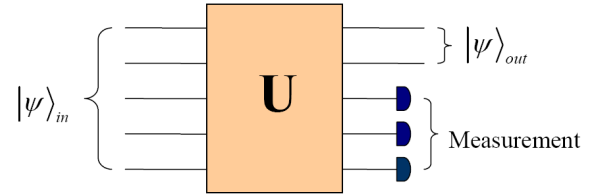}% Here is how to import EPS art
\caption{\label{fig:LOQSG} A Prototype of Linear Optical Quantum State Generator. It exploits linear operations, which eventually can be represented as a unitary transformation, and projective measurements to convert an input state into a target output state.}
\end{figure}

%\section{}%Formalism}
%section II: derive the formula for fidelity and success probability
Any LOQSG can be represented as in Fig.~\ref{fig:LOQSG}. Suppose we have $N$ input channels. They are composed with computational input states and ancilla channels. We want these inputs to be transformed to our expected output with projective measurement on the remaining ancilla ports. This process can be done by a linear optical device which we call a LOQSG \cite{VanMeter06}. This device is an $N$ dimensional unitary transformation. When a projective measurement determines a certain pattern of photons measured in some $M < N$ of the modes, it is considered successful, which leads to a preparation of the desired state in the remaining modes. Therefore, the device is probabilistic and it can fail in two aspects. Firstly, the measurement does not give out the expected pattern, which leads to the failed measurement. This can be improved by increasing the success probability of the device. Secondly, it does not provide the expected output state in the computational channels even when the measurement works perfectly. We note this kind of failure as fidelity less than unity. Since we can not measure the computational output during the computing process, we have to make sure that the fidelity is equal to one or numerically very very close to one.

The linear optical measurement-assisted transformation works as follows. We start from a computational input state $|\psi_{in}^C\rangle$ of $N-M$ modes, combined with ancilla state $|\psi_{in}^A\rangle$ in $M$ modes so that the input state $ |\Psi_{in} \rangle = |\psi_{in}^C\rangle \otimes |\psi_{in}^A\rangle $. The optical device induces a unitary transformation $\hat{U}$ of the $|\Psi_{in} \rangle$ state. After that a number-resolving photocounting measurement is applied to the $M$ ancilla modes. The latter is formally described by a Kraus POVM operator in ancilla modes $\hat{P} = |vacuum^A \rangle \langle k_{N-M+1}, k_{N-M+2}, ..., k_{N-M}| $. The resulting transformation of the computational state $| \psi_{in}^C \rangle $ is a contraction quantum map $|\psi_{out}^C \rangle = \hat{A} |\psi_{in}^C \rangle / \| \psi_{in}^C \| $ \cite{Kraus83}, where the action of the linear operator $\hat{A}$ is given by the following projection
\begin{equation}
\label{Eq:proj}
\hat{A} | \psi_{in}^C \rangle = \langle k_{N-M+1}, k_{N-M+2}, ..., k_{N-M}| \bm{\hat{U}} |\psi_{in}^C\rangle \otimes |\psi_{in}^A\rangle.
\end{equation}
In the context of the LOQSG problem, operator $\hat{A}$ contains all the information of state transformation.

The optical interferometer is considered formally  as canonical transformation of creation operators $a_i^{\dag} \rightarrow U_{ij} a_j^{\dag} $ induced by an $N \times N $ unitary matrix $U$. If the input state is given in the Fock representation as $|\Psi_{in} \rangle = |n_{1}, n_2, ..., n_{N-M}\rangle \otimes |n_{N-M+1},..., n_N \rangle$, the unitary transformation $\hat{U}$ in equation (\ref{Eq:proj}) is given by
\begin{equation}
\label{Eq:OutState}
|\Phi_{out}\rangle = \bm{\hat{U}}|\Psi_{in}\rangle = \prod_{i=1}^{N} \frac{1}{\sqrt{n_{i} !}} \left( \sum_{j=1} U_{i,j} a_{j}^{\dag} \right)^{n_i}  |0 \rangle.
\end{equation}
Transformation of Eq.~(\ref{Eq:OutState}) is a high-dimensional irreducible representation of the $N \times N$ matrix of the optical transformation $U$ ~\cite{Perelomov86}. In Fock representation, matrix elements of $\langle n | \hat{U} | n^{\prime} \rangle$ are calculated as permanents of matrix $U$ ~\cite{VanMeter06}.

Now we specify main properties of the Eq.~(\ref{Eq:proj}) relevant to numerical implementation of the optimization algorithm. In the computational Fock basis $| n^c \rangle$, the Eq.~(\ref{Eq:proj}) is described by matrix $ \hat{A}_{n_1^c, n_2^c} = \langle n_1^c | \hat{A} | n_2^c \rangle$, which is simply a submatrix of $\langle n | \hat{U} | n \rangle $. Thus $\hat{A}$ has a form of a set of polynomial functions in variables $u_{ij}$, computed using Eq.~(\ref{Eq:OutState}), so that Eq.~(\ref{Eq:OutState}) specifies explicit algebraic form of dependence of  $\hat{A}$ on  $\hat{U}$. If the total number of measured photons in ancilla modes $\sum_{i=N-M+1}^{N} k_i$ is the same as the number of input ancilla photons $\sum_{i=N-M+1}^{N} n_i$, then Eq.~(\ref{Eq:proj}) leaves the number of computational photons invariant. Since the dual-rail computational basis is just a subset of all possible states in the computational modes, the transformation matrix $\hat{A}_{n_1^c, n_2^c}$ is in general a non-square matrix, mapping the Hilbert space of the computation basis to a larger Hilbert space. For example, $\hat{A}_{n_1^c, n_2^c}$ for the CZ is a $10 \times 4$ matrix.

We now introduce the notion of operational fidelity of a transformation, which in general differs from the common measure of fidelity for a state transformation. From a physical point of view, the transformation $\hat{A}$ satisfies a $100 \% $  fidelity criteria if it is proportional to the target transformation operation $\hat{A}^T$, i.e., $\hat{A} \equiv \alpha \hat{A}^T$, where $\alpha$ is an arbitrary complex number. Since the target transformation is supposed to be a unitary gate, i.e., $\hat{A}_{Tar}^T \hat{A}_{Tar} = \hat{I}$. The operational fidelity condition also requires that desired transformation $\hat{A}$ satisfies operational unitarity condition $\hat{A}^T \hat{A} = S \hat{I}$, where $S= |\alpha|^2$ is the success probability of the transformation \cite{Lapaire03}. To formulate an algebraic estimate for the accuracy of the transformation, we consider complex rays $ \beta \hat{A} $ and $ \alpha \hat{A}^T $ as elements of complex projective space. The measure of closeness of elements in such a projective space is given by the Fubini-Study distance
\begin{equation}
\label{Eq:gamma}
\gamma (\hat{u}) = \arccos \sqrt{\frac{\langle \hat{A} | \hat{A}_{Tar} \rangle \langle \hat{A}_{Tar} | \hat{A} \rangle}{\langle \hat{A} | \hat{A} \rangle \langle \hat{A}_{Tar} | \hat{A}_{Tar}\rangle }},
\end{equation}
where the Hermitian inner product is $\langle A | B \rangle \equiv Tr(A B^{\dag} )/D_c$, and $D_c$ is the dimensionality of computational space.

In the numerical implementation of the optimization we used a nonsingular variable $F=\cos \gamma^2$, which we will refer to as fidelity in the rest of the paper.

Success probability $S$ of the transformation depends on the initial state $|\psi^c \rangle$ . The upper bound of $S$ is determined by the operator norm $\| \hat{A} \|^{Max} = Max (\langle \psi^c | \hat{A}^{\dag} \hat{A} | \psi^c \rangle / \langle \psi^c | \psi^c \rangle)$, and correspondingly the lower bound of $S$ is $\| \hat{A} \|^{Min} = Min  (\langle \psi^c | \hat{A}^{\dag} \hat{A} | \psi^c \rangle / \langle \psi^c | \psi^c \rangle)$. As a measurement of the success probability, we use the Hilbert-Schmidt norm $\| \hat{A} \|^{(HS)} = \sqrt{Tr(\hat{A}^{\dag} \hat{A})/D_c} $. It is easy to verify that $\| \hat{A}\|^{Min} \leq \| \hat{A}\|^{(HS)} \leq \| \hat{A}\|^{Max} $. As fidelity $F \rightarrow 1$, $ {\| \hat{A}\|^{Min}}/{\| \hat{A}\|^{Max}} \rightarrow 1$ and $S$ becomes a well defined parameter equal to $\| \hat{A}\|^{(HS)}$. We will refer to the Hilbert-Schmidt norm $\| \hat{A}\|^{(HS)}$ as success probability, keeping in mind that such a definition may not correspond to a success probability of transformation of specific state initial state.

As an example, consider first the nonlinear sigh (NS) gate. The NS gate with 2 ancilla modes is as follows
\begin{equation}
\label{Eq:NS}
\alpha | 010 \rangle + \beta | 110 \rangle + \gamma | 210 \rangle \longrightarrow \alpha | 010 \rangle + \beta | 110 \rangle - \gamma | 210 \rangle
\end{equation}
where $|\alpha|^2+|\beta|^2 + |\gamma|^2=1$. 
Because of conservation of the number of photons, the coefficient matrix $\hat{A}$ has a diagonal form ($3 \times 3$) with entries given explicitly as functions of $U_{n,m}$. 
\begin{eqnarray}
\label{Eq:NS_Coef}
|010\rangle: A_｛11｝ (\bm{U}) & = & U_{22}, \\
|110\rangle: A_｛22｝ (\bm{U}) & = & U_{12} U_{21}+U_{11} U_{22}, \\
|210\rangle: A_｛33｝ (\bm{U}) & = & U_{11}^2 U_{22}+2 U_{11} U_{12} U_{21}.
\end{eqnarray}
Then the operator success probability is
\begin{equation}
\label{Eq:NS_Succ}
%Succ Prob function for NS gate
S(\bm{U}) = \frac{1}{3} \sum_{i=1}^{3} | A_{ii} (\bm{U}) |^2.
\end{equation}
Since the target matrix $\hat{A}_{NS} = Diag(1,1,-1)$, the fidelity of $\hat{A}$ is trivially calculated as
\begin{equation}
\label{Eq:NS_Fid}
%fidelity function for NS gate
F(\hat{U}) = \frac{| A_{11} (\hat{U}) + A_{22} (\hat{U}) - A_{33} (\hat{U}) |^2 }{3 \Sigma_{i=1}^{3} | A_{ii} (\hat{U}) |^2},
\end{equation}

Similarly, The CZ gate, constructed with four ancilla modes, can be represented as
\begin{eqnarray}
\label{Eq:CZ}
\alpha | 1010 \rangle \otimes | 1010 \rangle  + \beta | 1001 \rangle \otimes | 1010 \rangle \nonumber \\
+ \gamma | 0110 \rangle \otimes | 1010 \rangle + \delta |0101 \rangle \otimes | 1010 \rangle \nonumber \\ 
\longrightarrow \alpha | 1010 \rangle \otimes | 1010 \rangle  + \beta | 1001 \rangle \otimes | 1010 \rangle \nonumber \\
+ \gamma | 0110 \rangle \otimes | 1010 \rangle - \delta |0101 \rangle \otimes | 1010 \rangle.
\end{eqnarray}
where $|\alpha|^2+|\beta|^2 + |\gamma|^2 +|\delta|^2=1$. In this case the matrix $\hat{A}$ is a $10 \times 4$ matrix, and the success probability has the form
\begin{equation}
\label{Eq:CZ_Succ}
%Succ Prob function for CZ gate
S(\hat{U}) = \frac{1}{4}\sum_{ij} |A_{ij}(\hat{U}) |^2,
\end{equation}
where i=1,...,4 and j=1,...,10. The corresponding fidelity function is
\begin{equation}                                                       
\label{Eq:CZ_Fid}
%Fidelity function for CZ gate
F(\hat{U}) = \frac{| A_{13}(\hat{U}) + A_{24}(\hat{U}) + A_{36}(\hat{U}) - A_{47}(\hat{U}) |^2}{4 \sum_{ij} | A_{ij}(\hat{U}) |^2},
\end{equation}
where $A_{13}$, $A_{24}$, $A_{36}$ and $A_{47}$ are coefficients of output states $|1010\rangle$, $|1001\rangle$, $|0110\rangle$ and $|0101\rangle$, respectively.

%\section{}%Annealing Genetic Algorithms with constraints}
%section III: Introduce the genetic algorithms and annealing GA with constraint 
If we write the unitary matrix as
\begin{equation}
\label{Eq:Generator}
U=U_0 \exp [\sum_{i=1}^{R} x_i g_i],
\end{equation}
where $x$ is an $R$ dimensional vector which identifies the unitary matrix $\hat{U}$ in an R-D unitary matrix space with $g_i$ as basis \cite{Nielsen00}. Each point of this space is a $N \times N$ unitary matrix. If we take Taylor expansion on the matrix exponential term, and truncate the polynomial with a proper error control, then we can represent the $\hat{U}$ as a polynomial function of $x$. Substituting it into the equations for success probability and fidelity, as derived in the last section, we have a polynomial function of $x$. Take $x$ as an individual or an abstract genotype in the language of genetics, we then have a formula ready for the genetic algorithm.  

We decided to explore genetic algorithms for the following reasons. Firstly, the problem is to optimize a multidimensional nonlinear function (eight modes for CZ \cite{CZVar}) with nonlinear constraints, i.e., a very large search space. In such cases there is no known traditional algorithm that has proven promising. (See Ref.~\cite{VanMeter06} for other approaches.) Genetic Algorithms (GAs) are therefore an attractive candidate. Secondly, the GAs can handle any form of function to be optimized, which will allow us to build a general scheme for our problem. In this paper, we focus on the NS and CZ gates, but the method can easily be adapted to any other LOQSG device. Thirdly, GAs are designed for searching the optimized results in a global space.

The first step in setting up a genetic algorithm is to choose a suitable fitness function. The simplest way is to select the fidelity given in Eqs.~(\ref{Eq:NS_Fid}) and (\ref{Eq:CZ_Fid}) as the the fitness function. Once we get a maximum fidelity, we can substitute the corresponding $x$ into the function to obtain the success probability, Eqs.~(\ref{Eq:NS_Succ}) and (\ref{Eq:CZ_Succ}). This approach is simple to implement but we only search the fidelity without worrying about the success probability. Since we know the fidelity has to be close to one to have a reliable result, we can consider fidelity as a constraint and success probability as the fitness function. Many computer scientists are working on constrained genetic algorithms \cite{coello02}. One of these is called the simulated annealing method \cite{constrainedGA96}. Using this method, we need to reformulate the fitness function so that it contains the constraint. The new fitness function can be written as
\begin{equation}
\label{Eq:Annealing}
\phi(x) = \alpha (F(x),T) S(x),
\end{equation}
with
\begin{equation}
\label{Eq:alpha}
\alpha (F(x),T) = e^{-(1-F(x))/T},
\end{equation}
where $F(x)$ is the fidelity function used as the constraint. The second parameter, referred to as the temperature $T$, is a function of the running time of the algorithm; $T$ tends to 0 (or very small values numerically) as execution proceeds. $ S(x)$ is the success probability, and  $\alpha (F(x),T)$ acts like a penalty so that the constraint can finally be satisfied. When the GA begins, we want the penalty to be small, i.e., $\alpha \approx 1$, so that the algorithm can search a bigger space to find the global maximum. When $T$ is large, which happens at the beginning of the execution, then $\alpha \approx 1$. As time goes on, $T \rightarrow 0$, then $\alpha \rightarrow 0$. It means the fitness tends to zero unless the constraint is satisfied, i.e., $F(\bm{x}) \approx 1$. Therefore, at the end of the GA run, we can get the optimized success probability with a fidelity of one. The details of this simulated annealing genetic algorithm are described in Ref.~\cite{constrainedGA96}. 

\begin{figure}
\includegraphics[width=8cm]{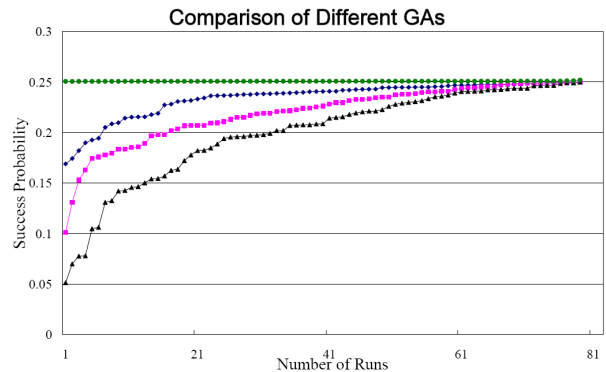} % Here is how to import EPS art Fig3.1 
\caption{\label{fig:NSCompare} The success probability for NS gate of different genetic algorithms. It shows the efficiency for different approaches: The black-triangle line indicates the result of a GA without any constraint, which takes fidelity as a fitness function, then calculates the success probability directly. The pink-square line describes the GA with a static-penalty when considering success probability as fitness function and fidelity as constraint, which is equivalent to setting the temperature in Eq.~(\ref{Eq:Annealing}) to be a small constant (e.g. $10^{-5}$). The blue-diamond line is the one which sets the annealing rate as $T(t)=-tan^{-1}(t)+\pi /2$. The green-dot line sets the annealing rate as $T(t)=1/\sqrt{t}$, where $t$ is time.}
\end{figure}
  
\begin{figure}
\includegraphics[width=8cm]{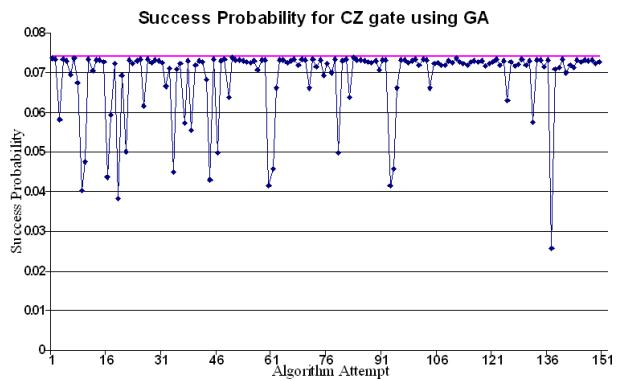} % Here is how to import EPS art Fig3.1 
\caption{\label{fig:CZ} The success probability for the CZ gate. Our goal is to verify and exceed if possible the known maximum success probability $2/27 \approx 0.074$ which is indicted by pink-dot line. The best result obtained so far is 0.0737 as indicted by blue-diamond line.}
\end{figure}

The annealing genetic algorithm provides a way to search the global maximum of success probability of a LOQSG system and guarantees that fidelity is equal to one at the same time. The disadvantage of this approach is that the efficiency delicately depends on the choice of the temperature annealing rate \cite{coello02}. In the following section, we investigate this problem using the NS gate as an example.

%\section{}%Simulation Results and Remarks}
%section IV: Results
%fig3.1 NS gate comparison
%fig3.2 CZ results

We use EOlib \cite{EO01} as the genetic algorithms framework. It provides a basic genetic operation template. In the case of the NS gate, we compared efficiency for different approaches. The comparison is shown in Fig.~\ref{fig:NSCompare}. The vertical axis denotes the success probability and the horizontal axis represents the events corresponding to each implementation. Each point indicates a complete run starting from a randomly selected population (a set of $x$. We rearrange the points in ascending order so that we can compare the efficiency easily. In Fig.~\ref{fig:NSCompare}, The black-triangle line indicates the result of a GA without any constraint, which takes fidelity as a fitness function, then calculates the success probability directly. The pink-square line describes the GA with a static-penalty when considering success probability as fitness function and fidelity as constraint, which is equivalent to setting the temperature in Eq.~(\ref{Eq:Annealing}) to be a small constant (e.g. $10^{-5}$). The third line with blue diamonds is the one which sets the annealing rate as $T(t)=-tan^{-1}(t)+\pi /2$. The fourth green line sets the annealing rate as $T(t)=1/\sqrt{t}$, where $t$ is time. From these plots, we can see that there are more chances for genetic algorithms with an annealing penalty to reach the global maximum of 1/4. The most impressive result is that when choosing the annealing rate as $1/\sqrt{t}$, the simulation can get the global maximum for each run. This is also consistent with the practical estimation in Ref.~\cite{constrainedGA96}.

Using this approach, we now design a CZ gate with high success probability. The results are shown in Fig.~\ref{fig:CZ}. They show a strong support for Knill's highest known success probability (2/27)\cite{KnillCZ02} being the actual maximum success probability. Due to the complexity of the CZ gate, it is not known if an analytical proof for determining the maximum success probability is possible. As discussed, once we get the $x$, we can construct the corresponding unitary matrix using Eq.~(\ref{Eq:Generator}). We can then discover the optimized design for the quantum circuit. 

In this work, we introduced simulated annealing genetic algorithms to look for an optimized linear optical quantum state generator. By investigating the NS gate, which has been theoretically proven to have a maximum success probability of 1/4, we found that we can reach the global maximum for each run if we carefully choose the annealing rate. We have found what we believe to be an upper bound for Success Probability for the CZ gate, our results showing a plateau around the best known published results of $2/27 \approx 0.074$ \cite{KnillCZ02}. Our best result so far having been 0.0737. Since this approach is focused on searching the unitary matrix space to optimize the quantum circuits, it can be generalized to other devices, for instance, to quantum sensors. 

\begin{acknowledgments}
%add acknowledgments here
We would like to acknowledge support from the Army Research Office and the Disruptive Technology Office.
\end{acknowledgments}
%\appendix
%\section{Introduction to Genetic Algorithms}
%end of Appendix


\begin{thebibliography}{15}
\bibitem[{CZV()}]{CZVar}
\bibinfo{note}{For an eight modes CZ gate, we need an $8 \times 8$ unitary
  matrix to execute the transformation. To generate such a complete matrix
  space, we need $8 \times 8-1=63$ dimensions in general, which is the
  dimension of $x$.}
\end{thebibliography}
\end{document}